\documentclass[showpacs,amsmath,amssymb, nobibnotes, aps, prl,showkeys]{revtex4-2}
\usepackage{graphicx}
\usepackage{dcolumn}
\usepackage{bm}
\usepackage{docs}
\usepackage{amssymb}
\usepackage{amsmath}
\usepackage{bigints}
\usepackage{relsize}
\expandafter\ifx\csname package@font\endcsname\relax\else
 \expandafter\expandafter
 \expandafter\usepackage
 \expandafter\expandafter
 \expandafter{\csname package@font\endcsname}%
\fi

\newcommand{\ltsima} {$\; \buildrel < \over \sim \;$}
\newcommand{\gtsima} {$\; \buildrel > \over \sim \;$}
\newcommand{\lta} {\lower.5ex\hbox{\ltsima}}
\newcommand{\gta} {\lower.5ex\hbox{\gtsima}}

\newcommand{\RNum}[1]{\uppercase\expandafter{\romannumeral #1\relax}}

\begin{document}

\title{Test of conformal theory of gravity as an alternative paradigm to dark matter hypothesis from gravitational
lensing studies}


\author {Shubhrangshu Ghosh$^{1}$, Mahasweta Bhattacharya$^{2}$, Yanzi Sherpa$^{1}$ and Arunava Bhadra$^{2}$  }

\affiliation{$^{1}$ Department of Physics, Shri Ramasamy Memorial (SRM) University, Sikkim, 5$^{\rm th}$ Mile Tadong, Gangtok, 737102, India. } 
\affiliation{$^{2}$ High Energy $\&$ Cosmic Ray Research Centre, University of North Bengal, Siliguri 734013, India.  }

\begin{abstract}
Weyl's conformal gravity theory, which is considered as a compelling alternative to general relativity theory, has been claimed to describe the observed flat rotation curve feature of spiral galaxies without the need of invoking dark matter. However, it is important to examine whether the Weyl theory can also explain the relevant gravitational lensing observations correctly without considering any dark matter. In this regard, the gravitational bending angle in static spherically space-time (Mannheim-Kazanas metric) in Weyl theory has been calculated by several authors over the last two decades, but the results are found largely divergent. In this work, we have revisited the problem and obtain the correct and consistent expression of the deflection angle in conformal gravity. Subsequently we perform the gravitational lensing analysis. We compare the prediction of Weyl gravity with the gravitational lensing observations of the rich galaxy clusters Abell 370 and Abell 2390 and is found that Weyl theory cannot describe the stated lensing observations without considering dark matter.        
\end{abstract}

\pacs{95.35.+d, 95.30.Sf, 04.50.Kd, 98.80.Es }
\keywords{Dark matter, relativity and gravitation, modified theories of gravity, observational cosmology}

\maketitle


\section{Introduction}

In recent years, fourth-order conformal (Weyl) gravity (a purely metric theory of gravitation based on the principle of conformal invariance), which has been suggested as a possible replacement to second-order theory of general relativity (GR), had generated considerable attention in the literature \cite{t1,t2,t3,t4,t5,t6}. The theory of Weyl gravity is particularly appealing, to the sense that this theory has been suggested to potentially resolve the problems of both DM and DE \cite{t5,t7}. Moreover, from the stand point of quantum gravity, the theory of weyl gravity enjoys the advantage of being conformally invariant and renormalizable (e.g., \cite{t8}). 

Conformal gravity theory has been found to be remarkably successful in fitting the empirical data of observed galactic rotational curves for a large sample of galaxies \cite{t5,t9,t10,t11,t12,t13}, thus eliminating the need to invoke any form of hypothetical dark matter (DM); the luminous mass component itself would then suffice. Conformal gravity theory has been demonstrated to be in agreement with the Tully-Fisher relation \cite{t13}, and can alternatively describe galactic halos without requiring unobservable DM \cite{t14}. It would then be worth exploring, whether, indeed, the theory of conformal Weyl gravity could be considered to be a reasonably viable  alternative to ``dark matter'' hypothesis ? A very compelling and significant consistency test in this regard would be to figure out, whether conformal gravity could account for the observed gravitational lensing in our Universe. Within the GR framework, the excess of lensing produced is attributed to the presence of DM. The conformal gravity model is thereby required to reproduce the observed gravitational lensing without the need to introduce any additional matter distribution, in consistency with its description of observed rotation curves of galaxies. In the present article, we seek to address this issue from the stand point of gravitational lensing measurements. 

Since its discovery in the late nineteen-seventies, gravitational lensing phenomenon has gradually emerged to become a prominent observational tool in astronomy and cosmology. The lensing measurements have been identified as distinctive tool to address the mass discrepancies in the Universe. It has been widely argued that observations combining galactic rotation curves and gravitational lensing can furnish better insight about the gravitational field and the sub-structure within galactic halos \cite{t15,t16}. In addition, lensing measurements can be an effective tool to possibly discriminate between DM hypothesis and alternative models to DM. In recent years, several authors have performed lensing study for background non-Schwarzschild space-time geometries 
(e.g., \cite{t17,t18,t19,t20,t21}). Considering  Weyl gravity as an alternative paradigm to DM hypothesis, we seek to perform lensing studies under this theory and compare them with observations, under weak lensing scenario. 

Since conformal Weyl gravity has been revisited by Mannheim and Kazansas (1989) \cite{t5}, several authors in recent times have obtained expressions for deflection of light and performed lensing analysis in conformal gravity in Mannheim-Kazansas space-times (e.g., \cite{t22,t23,t24,t25,t26,t27,t28,t29}), particularly in the context of weak lensing scenario. But the bending angle expressions given by the different authors are found different and non-convergent. Our first job will be to critically review the concerned works on gravitational bending angle in Weyl gravity and to evaluate the right expression of the bending angle. 

The rest of the paper is organized as follows: In next section, we furnish the expressions for light deflection under conformal gravity and obtain suitable expression for gravitational lensing, relevant to our purpose. In \S 3, we make a comparison of our theoretical results with observed data to check whether our theoretical findings remain consistent with observed gravitational lensing measurements. Finally, we culminate in \S 4 with a summary and discussion. 

\section{Gravitational bending angle in Mannheim-Kazansas space-time} 

The conformally invariant Weyl action is given by

\begin{eqnarray}
I= \alpha \int  d^4x \sqrt{-g} C_{\lambda\mu\nu\kappa} C^{\lambda\mu\nu\kappa} \, , 
\label{1}
\end{eqnarray}

where $C_{\lambda\mu\nu\kappa}$ is the Weyl tensor and $\alpha$ is dimensionless constant. The static spherically symmetric exterior solution of Weyl gravity is obtained by Mannheim and 
Kazanas \cite{t5} [hereinafter MK], which is given by

\begin{eqnarray}
ds^2=  - {\mathcal{B}}(r) dt^2 + \frac{dr^2}{ {\mathcal{B}}(r)  } + r^2 \left(d\theta^2 + \sin^2\theta \, d\phi^2  \right) \, , 
\label{2}
\end{eqnarray}
where, ${\mathcal{B}}(r) $  is given by 

\begin{eqnarray}
{\mathcal{B}}(r) =  \left(1 - 6 m \gamma \right)^{1/2}  - \frac{2 m}{r} + \gamma r - k r^2 \, , 
\label{3}
\end{eqnarray}
where, $m = \frac{GM}{c^{2}} $, $M$ is the usual luminous mass, $\gamma$ and $k$ are integration constants which could be appropriately determined by matching the theoretical predictions with observations.  
$\gamma$ is associated with the inverse Hubble length, with $\gamma \simeq \frac{1}{R_H}$, and $k$ is related to Cosmological constant, with $k = \frac{\Lambda}{3}$. Since $k$ is very small and our objective is to find out the effect of MK parameter on gravitational bending angle, we shall neglect the effect of $k$ in this work. 
The motion of a time-like particle in MK space-time can be obtained from the geodesic equations. The photon trajectory in MK space-time on the equatorial plane is given by

\begin{eqnarray}
\frac{1}{r^2}\left(\frac{dr}{d\varphi}\right)^2 + \mathcal{B} -\frac{r^2}{b^2} = 0 \, , 
\label{4}
\end{eqnarray}

where $b$ is a constant of integration, which can be replaced by the distance of closest approach ($R$) where $dr/d\varphi$ vanishes. For convenience one defines $u=1/r$. In terms of the new variable u, the above equation reduces to 
\begin{eqnarray}
\frac{d^{2}u}{d\varphi ^{2}}=-\frac{\gamma }{2} -\alpha u+3mu^{2}.
\label{5}
\end{eqnarray}
where $\alpha =(1-6m\gamma )^{1/2}$. 

The gravitational deflection angle in Weyl gravity was first estimated by Edery and Paranjape \cite{t22} using the MK metric under weak field and small-bending-angle approximations. They calculated the change in the coordinate angle under geodesic motion by integrating the above equation with respect to $r$ following the conventional method of gravitational deflection angle determination. They found the following expression for the deflection angle ($\hat{\alpha}$) in the leading order of $m$ and $\gamma$

\begin{eqnarray}
\hat{\alpha} \equiv 2 \psi \, \simeq \, \frac{4m}{R} - \gamma R  \, ,
\label{6}
\end{eqnarray}

where $R$ is the distance of closest approach, which suggests that the MK parameter contribution to bending angle is negative: it diminishes the gravitational mass (of the lens object) led bending angle which is contradictory to the observations. The sign of MK parameter $\gamma$ has to be the opposite of the value obtained by fitting the galactic rotation curve in order to cause a stronger deflection than the GR prediction. An \textit{ad hoc} way to resolve the disparity is by selecting a proper gauge for the metric before computing the deflection angle \cite{t30}. 

An important feature of the MK space-time metric is that it is asymptotically non-flat in contrast to Schwarzschild space-time of GR, which becomes Minkowski space-time at large distances. Rindler and Ishak \cite{t31} had shown, in the context of Kottler space-time, that the conventional approach is not a proper method for determining bending angle in asymptotically non-flat space-time because the limit $r \rightarrow \infty$ makes no sense for such geometries. Rindler and Ishak proposed an invariant angle method, in which essentially the angles that the tangent of the light trajectory makes with respect to a reference direction at the source and the observer positions are estimated. Sultana and Kazanas \cite{t23} and Bhattacharya et al. \cite {t24} recalculated the bending angle in MK space-time under weak field and small-bending-angle considerations adopting Rindler-Ishak method. Sultana and Kazanas got the following expression for total bending 
angle \cite{t23}

\begin{eqnarray}
\hat{\alpha} \equiv 2 \psi \, \simeq \, \frac{4m}{R} - \frac{2m^2\gamma}{ R} \, .
\label{7}
\end{eqnarray}

The second term in the right-hand side is though negative but insignificant in comparison to the first term (Schwarzschild deflection angle). Bhattacharya et al. \cite{t24}, however, got the same $-\gamma R$ term in the leading order as obtained in \cite{t22}. Later, Cattani et al. \cite{t27} argued that when all the terms in leading order of $\gamma$ and $m$ are properly taken into consideration, the total bending angle formula will become 

\begin{eqnarray}
\hat{\alpha} \equiv 2 \psi \, \simeq \, \frac{4m}{R} + \frac{15m^2\gamma}{ R} \, .
\label{8}
\end{eqnarray}

Importantly, the second term is positive, though still insignificant compare to the first term. In another work Sultana \cite{t26} also obtained similar expression with positive contribution of $\gamma$ but got the coefficient of $\frac{m^2\gamma}{ R}$ as $`6'$ instead of $`15'$ as in the above equation. Lim and Wang tackled the problem of evaluating bending angle in slightly different 
way \cite{t28}. They first obtained the exact bending angle expression without any approximation in terms of various trigonometric and elliptic functions, series expanded it and considered only the leading order terms in $m$ and $w$ 
$\left[w\equiv \gamma R/(m/R) \right]$. Their approximate total bending angle expression reads \cite{t28} [see Eq. (35) of Lim and Wang]
\begin{eqnarray}
\begin{split}
\hat{\alpha} \equiv  2 \psi \, \simeq  \  & \frac{4 m}{R} \, + \, \left( -4 + \frac{15}{4} \pi \right) \frac{m^2}{R^2} \, +  \,  \left( \frac{98}{3} - \frac{15}{2} \pi \right) \frac{m^3}{R^3} \, + \, \left[ \frac{2 m^2}{R^2} +  \left( -4 + \frac{15}{4} \pi \right)  \frac{m^3}{R^3}   \right] w \, \\
& + \, \left[- \frac{m}{8 R}   - \left(\frac{3}{8} - \frac{15}{128} \pi \right)  \frac{m^2}{R^2}   -  \left(- \frac{5}{16} - \frac{15}{64} \pi + \frac{225}{2048} \pi^2 \right)  \right] w^2 
\, + \,  \mathcal{O}({m^4}/{R^4}, w^3)  \, 
\end{split}
\label{9}
\end{eqnarray}

It is noticed that in contrast to the previous works, there is an additional term $2m\gamma$ in Eq. (8) which is significantly larger than the leading correction term ${15m^2\gamma}/{ R}$ in Eq. (7). However, the bending angle expression of Lim and Wang [Eq. (8)] is not valid at large distance scales. This is because the expansion was made in power of $w$ under the assumption that $w < 1$. But at the outer part of a galaxy $\gamma R$ has to be larger than $m/R$ in order to explain the flat rotation curve features which implies $w>1$ at such distance scales. The Lim-Wang bending angle diverges for $w>1$ as revealed from the first term in $w^2$ expansion. 

Under the present circumstances, it then becomes necessary to re-investigate the gravitational deflection angle in MK spacetime.   
Defining $\alpha u =\overline{u}$, $\alpha^{1/2} \varphi = \overline{\varphi}$, the geodesic equation (5) up to the second order in $m$ takes the form
\begin{eqnarray}
\frac{d^{2}\overline{u}}{d\overline{\varphi} ^{2}}=-\frac{\gamma }{2} -\overline{u}+3m\overline{u}^{2}.
\label{10}
\end{eqnarray}
Following the standard perturbation approach, the solution of the above equation up to second order accuracy in $m$ and $\gamma$ reads 
\begin{eqnarray}
\overline{u} &=& -\frac{\gamma}{2} + u_o cos\overline{\varphi} +\frac{3m\gamma u_o}{4}\left(cos\overline{\varphi}-\overline{\varphi}sin\overline{\varphi} \right)  \nonumber \\
&& + \frac{3 m^2 u_o^3}{16} \left(20\overline{\varphi}sin\overline{\varphi}+cos3\overline{\varphi} \right),
\label{11}
\end{eqnarray}%
where $u_o$ is a constant, which can be eliminated in terms of distance of closest approach. At the distance of closest approach ($r=R=1/u_m$) $\frac{d\overline{u}}{d\overline{\varphi}} =0$. One then gets
\begin{eqnarray}
\frac{d\overline{u}}{d\overline{\varphi}} &=& - u_o sin\overline{\varphi} + mu_o^2 sin 2\overline{\varphi} + \frac{3m\gamma u_o}{4}\left(2sin\overline{\varphi}-\overline{\varphi}cos\overline{\varphi} \right)  \nonumber \\
&& + \frac{3 m^2 u_o^3}{16} \left(20 sin\overline{\varphi}+ 20 \overline{\varphi} cos\overline{\varphi} -3 sin 3\overline{\varphi} \right),
\label{12}
\end{eqnarray}

The above term $\frac{d\overline{u}}{d\overline{\varphi}}$ vanishes when $\overline{\varphi}=0$. Consequently, one gets,

\begin{eqnarray}
\frac{1}{R}=u_m =-\frac{\gamma }{2} + u_o -\frac{3 m\gamma u_o}{4} + m u_o^2 + \frac{3 m^2 u_o^3}{16}
\label{13}
\end{eqnarray}

The conventional approach of evaluating light deflection angle is not applicable for asymptotically non-flat spacetimes like the MK metric, one may instead apply the Rindler-Ishak method of invariant angle, according to which the deflection angle (for trajectory one side of the lens) is given by

\begin{eqnarray}
\tan \psi =\frac{\mathcal{B}^{1/2}r}{\left\vert \mathcal{A}\right\vert },
\label{14}
\end{eqnarray}
where $\mathcal{A}(r,\varphi )=\frac{dr}{d\varphi }$. We calculate the deflection angle considering the position of the source, which is far away from the lens, is at $\varphi = \pi/2$. Using 
Eqs. (12), (13), and (14), we obtain the appropriate deflection angle for the total trajectory up to the second order accuracy in $m$ and $\gamma$ 

\begin{eqnarray}
2\tan \psi \sim 2\psi \simeq \, \frac{4m}{R} +  \left(2+\frac{3\pi}{4}\right) m\gamma + \left(\frac{15 \pi}{4} -4\right) \frac{m^2}{R^2}  \equiv \hat{\alpha} \, .
\label{15}
\end{eqnarray}

One should note that the effect of higher order terms to the total bending angle is negligible, and hence retaining to second order accuracy is pertinent for our purpose. We notice that our above derived expression differs from that given by Cattani et al. \cite{t27} and also the expression of Sultana \cite{t26}. Such a discrepancy appears because Cattani et al. \cite{t27} used the relation between the distance of closest approach and the integration constant (say $r_o$) of the path equation $\left( \frac{1}{R} = \frac{1}{r_o} + \frac{m^2}{r_o^2} \right)$ which is valid for Schwarzschild metric but modifies in presence of $\gamma$. When the correct expression of the distance of closest approach is used, one would get the $2m\gamma$ term in the expression for the bending angle. 
Sultana \cite{t26} used the correct expression of the distance of closest approach but later ignore $\gamma$ from that relation, citing it is too small. 

It is interesting to mention that recently, Kasikci and Deliduman (2019) \cite{t29} recalculated the deflection angle in space-time, which is conformally equivalent to the MK space-time. Since null geodesic remains the same in any conformally equivalent space-time, Kasikci-Deliduman bending angle formulation is valid for the MK space-time. They choose a conformally related coordinate system of the usual coordinate system of MK space-time, so that there is no cosmological horizon in the transformed metric in the new coordinate system. Consequently, Kasikci and Deliduman \cite{t29} employ the conventional approach of bending angle calculation without bothering the Rindler-Ishak technique. They 
got the following expression for total bending angle, which reads

\begin{eqnarray}
\hat{\alpha} \equiv  2 \psi \, \simeq  \, \frac{4m}{r_o} + 2m\gamma + \frac{m^2}{ r_o^2} \left(1+\gamma r_o\right) \left(\frac{15 \pi}{4} - 4\right) \, . 
\label{16}
\end{eqnarray}

From the above expression, it seems to appear that the authors retained the expression up to the third order accuracy with the term $\gamma m^2$, however, other possible third order terms do not explicitly appear in the above bending angle expression. Nonetheless, Eq. (15) renders an appropriate and consistent expression for total bending angle till second order accuracy. Here it is interesting to observe that, to second order accuracy, Kasikci-Deliduman bending angle formula is close to our bending angle formula expressed in Eq. (15), with a slight difference in the coefficient of $m \gamma$. Appropriate for our purpose, we shall use the correct and consistent bending angle expression as furnished in Eq. (15) for the gravitational lensing studies as reported below.

 
\subsection{Lensing analysis under conformal Weyl gravity} 

In the limit of thin lens approximation, appropriate for astronomical objects, the gravitational lens equation in the weak lensing scenario is given by  (see \cite{t20,t27,t32})

\begin{eqnarray}
\beta = \theta  \, - \, 2 \psi \, \frac{D_{ls}}{D_{os}} \, , 
\label{17}
\end{eqnarray}
where $\beta$ denotes the angular source position, $\theta$ denotes the angular position of the images, $D_{ls}$ is the distance from the lens to the source, and $D_{os}$ denotes the distance from the observer to the source. Note that the above expression in Eq. (17) is obtained considering asymptotic flatness, as a first approximation \cite{t20}. When the source, lens, and the observer are all aligned in one direction, i.e., they are all lined up along optical axis, one have $\beta = 0$. The original light source would then produce a ring around the lens, which is what is known as Einstein ring. 
From Eq. (17), the angular radius of the Einstein ring ($\theta_E$) would then be defined as  

\begin{eqnarray}
\theta_E  \, = \, 2 \psi \, \frac{D_{ls}}{D_{os}} \, . 
\label{18}
\end{eqnarray}
If ``$b$'' is the impact parameter, one can express ${R} \sim b$. In small angle approximation, it then yields 

\begin{eqnarray}
b = \theta_E  \, D_{ol} \, , 
\label{19}
\end{eqnarray}
where, $D_{ol}$ denotes the distance of the lens from the observer. As we are performing our lensing analysis in Weyl conformal gravity, in the present context, the radius of the Einstein ring $\theta_E$ can also be denoted as ``Weyl angle $\left(\theta_{\rm Weyl}\right)$'' \cite{t27}. Fitting the rotation curves of the galaxies, one obtains the value of $\gamma$ which 
reads \cite{t13,t33}   $\gamma = 5.42 \times {10}^{-41} \, {\rm cm}^{-1} \, \frac{M}{M_{\odot}} + 3.06 \times {10}^{-30} \, {\rm cm}^{-1} $. For simplicity, we denote 
$a_1 = 5.42 \times {10}^{ -41} \, {\rm cm}^{-1}$, and $b_1 = 3.06 \times {10}^{ -30} \, {\rm cm}^{-1}$. Expressing $R$ in terms of $b$, using Eq. (19), and substituting the expression for bending angle furnished in Eq. (15) in Eq. (18), the corresponding lens equation relevant for our purpose in terms of luminous mass $M$ in conformal (Weyl) gravity is as follows:

\begin{eqnarray}
\begin{split}
\theta^2_{\rm Weyl} \, \frac{D_{os} \, D_{ol}}{D_{ls}}\, \simeq \, & \,  r_s \, \left[\left(1 + \frac{3}{8} \pi \right) \, a_1 \, \theta_{\rm Weyl} \, D_{ol} \,  + \, \left(\frac{15}{16} \pi  - 1\right) \, \left(\frac{1}{\theta_{\rm Weyl} \, D_{ol}} \right)  \, r_s \right] \, \left(\frac{M}{M_{\odot}}\right)^2 \,  \\
& + \, r_s \, \left[2  +  \left(1 + \frac{3}{8} \pi \right) \, b_1 \, \theta_{\rm Weyl} \, D_{ol} \right] \, \frac{M}{M_{\odot}} \, ,
\end{split}
\label{20}
\end{eqnarray} 
where, $r_s = {2GM_{\odot}}/{c^2}$ is the Schwarzschild radius of the Sun. The above expression in Eq. (20) has been furnished using the value of 
$\gamma$ in terms of $a_1$ and $b_1$.  

Considering conformal Weyl gravity to be a suitable substitute to DM hypothesis, the ``Weyl angle $\left(\theta_{\rm Weyl}\right)$'' would then be equivalent of $\theta_E$, the usual Einstein radius of the lens system (i.e., $\theta_{\rm Weyl} \equiv \theta_E$ ). Note that, for any gravitational lens system, in principle, $\theta_E$,  $ D_{ls} $, $D_{os}$, $D_{ol}$ are observationally measurable quantities. Corresponding to the known values of these four quantities, one can then estimate the luminous mass $M$ of the lens system in Weyl gravity. Our objective here, is to compare the 
theoretically estimated value of luminous mass in conformal gravity, to that of estimated total mass ($M_T$) of observed gravitational lens systems. For our purpose, here, we consider the scenario 
for gravitational lensing by galaxy clusters Abell 370 and Abell 2390. It is worth mentioning here, that for galaxy–galaxy weak gravitational lensing, the analysis is based on distortion of the shape of images of distant background galaxies on account of lensing by foreground galaxies. However, the extent of distortion of images in this context is significantly less, unless one has massive lensing object. Galaxy-galaxy weak lensing technique, thereby, may not necessarily be adequate enough to obtain pertinent properties of individual galaxies. Hence, for our purpose, it would be more 
desirable to consider lensing by galaxy clusters, rather galaxy-galaxy weak lensing. For more details regarding this, readers may look into the following Ref. \cite{t20}. 

\section{Comparison of our theoretical findings with that of the lensing observations due to Abell 370 and Abell 2390}

Abell 370 is a well studied gravitational lens system. This rich galaxy cluster comprising of several hundreds of galaxies is found to be located  at a distance of 1464 megaparsecs (Mpc) away from us in the constellation Cetus. Extensive spectro-photometric studies of Abell 370 were conducted, and ``giant luminous arcs'' were first detected at redshift $z_l = 0.374$ \cite{t34,t35}. The longest arc `A0' has been observed at redshift $z_s = 0.724$. From a thorough analysis it has been suggested that the observed luminous arcs are gravitationally lensed images of distant background galaxies \cite{t36,t37,t38,t39} . Through the gravitational lens modeling the total mass ($M_T$) of the galaxy cluster Abell 370 could then be estimated using the lensed images. Our objective here is to measure the total mass of the lensing cluster Abell 370 from one of the giant luminous arcs. For this, we consider the longest arc `A0' having a radius of curvature of around  $25''$ and treats it as an Einstein ring \cite{t36,t38}. Usually, galaxy clusters consist of a complex matter distribution, and may not be appropriate to consider them as either point masses or having spherically symmetric mass distribution. Nonetheless, as a first approximation, spherically symmetric lens model could still be employed to obtain the results within same order of magnitude as a more realistic case by examining the giant luminous arcs observed in clusters \cite{t39,t40}. If source, lens and observer are all aligned along the optic axis, modeling the lens as purely Schwarzschild space-time, one obtains the net mass $M_T$ at the zeroth-order accuracy within the Einstein radius ($\theta_E$), given through the relation (see e.g., \cite{t32}). 

\begin{eqnarray}
\frac{M_T}{M_{\odot}} =  \theta^2_{E} \, \frac{1}{2 r_s} \, \frac{D_{os} \, D_{ol}}{D_{ls}} \, , 
\label{21}
\end{eqnarray} 
where, $D_{ls}$, $D_{os}$, and $D_{ol}$, here, are angular distances. Using the observed redshift values of $z_l$ and $z_s$, one can determine the distances of the lens ($D_{ol}$) and 
background galaxy ($D_{os}$). The angular distance $D_{ls}$ can be obtained using the red shift value $z_{ls} = 0.255$. A concordance cosmological model 
of ($ \Omega_m; \Omega_{\Lambda}; \Omega_k $) = ($0.286; 0.714; 0$) with the value of Hubble constant $H_o = 69.6$ is used to estimate the distances from redshifts 
of lens and source. The angular distances have been estimated to be ($D_{ol} = 1073.2 \, {\rm Mpc}; D_{os} = 1512.1 \, {\rm Mpc}; D_{ls} = 825.2 \, {\rm Mpc}$). 
We now wish to compare the value of $M_T$ obtained from Eq. (21) with that of luminous mass $M$ using Eq. (20) in Weyl gravity, corresponding to Abell 370. 
Substituting the values of radius of Einstein's ring ($\theta_E = 25''$), and angular distances ($D_{ls}$, $D_{os}$, and $D_{ol}$) corresponding to Abell 370, the lens equations in Weyl gravity in terms of luminous mass $M$ for our case [given through Eq. (20)] can be re-written as \\ 

{\large For Abell 370} \\ 
\begin{eqnarray}
\left(\frac{M}{M_{\odot}} \right)^2 \, + \,  4.095928 \times 10^{16} \left(\frac{M}{M_{\odot}} \right) \, - \, 6.181836 \times 10^{30} = 0 \, . 
\label{22}
\end{eqnarray}
The photometric measurements reveal that the mass-to-light ratio within the arc `A0' in Abell 370 is about $\sim 100$ or even around $\sim 190$ \cite{t35,t36,t38}. This suggest that the luminous mass ($M$) in this galaxy cluster should be at least an order of magnitude smaller than the total mass ($M_T$) of the cluster. 

Let us now focus on other lensing cluster Abell 2390 that we considered here. This galaxy cluster is observed at redshift ($z_l = 0.231$) \cite{t40}. The spectro-photometric observations display a very linear arc observed at redshift $z_l = 0.913$, with a radius of curvature of around $37''$ \cite{t40,t41}. As in the previous case for Abell 370, we treat this arc as an Einstein ring of radius ($\theta_E = 37''$). Using the observed redshift values and the concordance cosmological model (furnished before), one obtains the value of angular distances ($D_{ol} = 766.6 \, {\rm Mpc}; D_{os} = 1635.6 \, {\rm Mpc}$), pertaining to Abell 2390. The corresponding value of angular distance $D_{ls}$ can be obtained using the red shift value $z_{ls} = 0.255$. Modeling the lens as Schwarzschild space-time, one can then obtain the total mass $M_T$ within the Einstein radius ($\theta_E$) from Eq. (21), corresponding to the lensing cluster Abell 2390. In a similar fashion as in the previous case, substituting these values of $\theta_E$ and distances ($D_{ls}$, $D_{os}$, and $D_{ol}$), the lensing equation in conformal (Weyl) gravity in terms of luminous mass $M$ for Abel 2390 can be re-written as \\

{\large For Abell 2390} \\ 
\begin{eqnarray}
\left(\frac{M}{M_{\odot}} \right)^2 \, + \,  3.88638 \times 10^{16} \left(\frac{M}{M_{\odot}} \right) \, - \, 6.104065 \times 10^{30} = 0 \, . 
\label{23}
\end{eqnarray}
Again, the observed mass-to-light ratio indicates that the luminous mass in this cluster be at least an order of magnitude less than the total mass ($M_T$) of the cluster [96] \cite{t42}. 

In Table 1, we depicted the values of total mass $M_T$ estimated from lensing data using Eq. (21), as well the luminous mass $M$ in conformal Weyl gravity from Eqs. [(22) - (23)], corresponding to galaxy clusters Abell 370 and Abell 2390, respectively. Our results clearly reveal that for both these galaxy clusters, theoretically estimated luminous mass considering the lensing in conformal gravity theory remains almost identical to the mass measured from observed gravitational lensing in purely Schwarzschild space-time. This is not what is being expected, as the luminous mass in a galaxy cluster should be at least $1 - 2$ orders of magnitude less than the total mass of a cluster, based on photometric measurements. 

Here it is worth mentioning that for Abell 370, the estimated total mass considering the lens as Schwarzschild space-time from our simplified model [Eq. (21)] is consistent and is of the same order of magnitude with that of the other measured values of mass available in the literature; see for e.g., in references \cite{t38,t39}, from Subaru weak lensing measurements \cite{t43} and from the hubble space telescope (HST) observations \cite{t44,t45}. However, note that, for Abell 2390, enough relevant information is not available in the literature in this regard. For more details regarding the observational data corresponding to the lensing clusters Abel 370 and Abel 2390, readers are advised to see Refs. \cite{t20,t32}. 

Our objective in this article is to examine whether the non-baryonic part of the total mass can be alternatively described by conformal gravity theory. Although the data overwhelmingly supports that 
the theory of conformal gravity can adequately produce the effective potential consistent with the observed rotational curves of the galaxies without the need to invoke DM, however, this theory seems to be inconsistent in explaining gravitational lensing measurements (at least corresponding to the galaxy clusters we considered here). Assuming that the conformal Weyl gravity be rightly chosen to be an alternative paradigm to DM hypothesis, consistent with the lensing measurements, the expected value of the luminous mass estimated in Weyl gravity should then be of much lesser value than that obtained from our analysis; the contribution of the $\gamma$ term in the metric should then alternatively account for the excess mass discrepancy of the lensing object. However, this is contrary to our findings. This seems to be in sharp contrast to the GR scenario (i.e., for purely Schwarzschild geometry) where the estimated total mass of a cluster is the aggregate of the luminous and hypothetical non-baryonic DM.

\begin{table}
\large
\centerline{\large Table 1} 
\centerline{\large Estimated total and luminous mass of Abell 370 and Abell 2390} 
\begin{center}
\begin{tabular}{cccccccccccc} \\
\hline
\hline
\noalign{\vskip 2mm}
${\rm Lensing}$ & $z_l$ & $z_l$ & $z_{ls}$ & $\theta_E$  & ${M_T}/{M_\odot}$  & ${M}/{M_\odot} $ \\
$ {\rm object}$ & $ $ & $ $ & $ $ & $({\rm in \, arc \, secs})$ &  $(\rm from \, lensing $&  $[{\rm luminous \,  mass} $ \\
$ $ & $ $ & $ $ & $ $ & $ $ & ${\rm observation}) $ &  ${\rm from \, Eq. (22)  } $  \\
$ $ & $ $ & $ $ & $ $ & $ $ & $ $ &  ${\rm and \, Eq. (23) }] $  \\
\hline
\hline
\noalign{\vskip 2mm}
${\rm Abel \, \, 370 }$  &  ${\rm 0.374}$ & ${\rm 0.724}$  & ${\rm 0.255}$  & ${\rm 25}$ & $\sim {\rm 1.51 \times 10^{14}}$ &  $\sim {\rm 1.504 \times 10^{14}} $ \\
\noalign{\vskip 2mm}
${\rm Abel \, \, 2390 }$  &  ${\rm 0.231}$  & ${\rm 0.913}$ & ${\rm 0.554} $ & ${\rm 37} $ & $\sim {\rm 1.57 \times 10^{14}}$ & $\sim {\rm 1.564 \times 10^{14}} $ \\
\noalign{\vskip 2mm}
\hline
\hline
\end{tabular}
\end{center}
\end{table}

\section{Discussions}

In this article we intended to test whether conformal (Weyl) gravity could be treated as a credible alternative to `dark matter' hypothesis. There have been considerable argument in the literature that conformal gravity theory could circumvent the need to introduce any form of non-baryonic dark matter, owing to the fact that this theory could effectively describe the observed flatness in the galactic rotation curves. A pertinent question that arises is whether the theory of conformal gravity could then adequately account for other astrophysical and/or cosmological observations in our Universe, like observed gravitational lensing. 

Several lensing observations (see the Refs. in \S 1, and previous sections in this article) had already indicated the presence of an unseen added mass (at least with an magnitude of order larger) over the luminous mass component of the gravitationally-lensed objects, particularly when one considers galaxy clusters as lenses. Within the customary GR framework, the (large) extra lensing produced is attributed to the presence of 
dark matter. It then entails that the observed deflection of light results due to total matter 
distribution comprising of hypothetical `dark matter' in addition to luminous mass component. In contrast, in conformal gravity theory, this hypothesis is avoidable. If, indeed, conformal gravity model be considered to be a plausible substitute to DM hypothesis, this model is then be required to reproduce the observed gravitational lensing, in consistency with its description of observed flatness in galactic rotation curves. More precisely, within the paradigm of Weyl gravity, the excess of lensing produced should arise due to the contribution of the $\gamma$ term in the metric, thereby, eliminating the need to invoke any hypothetical dark matter. In the present article, our objective was to examine this issue, by having a consistency check with the gravitational lensing data.

In this regard, it is important to note that the gravitational bending angle in static spherically space-time (Mannheim-Kazanas metric) in Weyl theory has been calculated by several authors over the last two decades, but the results are found largely divergent, and grossly inconsistent. In the present work, we have revisited the problem and obtain the correct and consistent expression of the deflection angle in conformal gravity, appropriate to second order accuracy. For our purpose, we have considered lensed galaxy clusters; a well studied gravitationally lensed rich clusters Abel 370 and Abel 2390. Using the lensing data, and modeling the lens as a purely Schwarzschild space-time (i.e., for usual GR case), we obtained the total mass of the cluster $M_T$ at the zeroth-order accuracy, for the two stated galaxy clusters. On the other hand, in context of our lensing analysis corresponding to our conformal (Weyl) gravity, we chose our appropriate expression for total light bending angle expressed through Eq. (15). Considering small angle approximation and to leading order in $\gamma$ and mass term, we estimated the luminous mass $M$ corresponding to clusters Abel 370 and Abel 2390 using lensing data, in Weyl gravity. Our findings show that for both the lensed galaxy clusters considered here, the estimated luminous mass in conformal gravity remains almost identical to that of the estimated total mass of the clusters. However, this should not be the case as the luminous mass in a galaxy cluster should be at least an order less than the total mass of a cluster; in conformal gravity, the additional lensing produced is supposed to occur due to the contribution of the $\gamma$ term in the metric.  

Our study indicates that conformal gravity theory seems to be inconsistent in explaining gravitational lensing measurements (at least corresponding to the galaxy clusters we considered here). A similar conclusion was reached in \cite{t46} using the (improper) expressions of \cite{t23} and \cite{t28}. Although this model is quite effective to describe observed flatness in galactic rotation curve without the need to invoke any dark matter, however, it fails to reproduce observed gravitational lensing (within an acceptable limit). Conformal (Weyl) gravity, thus, may not be a credible alternative to `dark matter hypothesis'. Here it might be important to note that, although modified theories of gravity like conformal gravity can conceivably describe a part of observations without the need to introduce non-baryonic DM, however the theory seems to lack adequacy to conform with other relevant astrophysical/cosmological observations. 

\section*{Acknowledgments}
The work is supported by Science and Engineering Research Board (SERB), DST through grant number CRG/2019/004944.



\end{document}